# Photoswitchable Single-Walled Carbon Nanotubes for Super-Resolution Microscopy in the Near-Infrared

Antoine G. Godin[1,2,3,4], Antonio Setaro[5], Morgane Gandil[1,2,†], Rainer Haag[6], Mohsen Adeli[6,7], Stephanie Reich[5] & Laurent Cognet[1,2,*]

[1] *Univ. Bordeaux, Laboratoire Photonique Numérique et Nanosciences, UMR 5298, F-33400 Talence, France*
[2] *Institut d'Optique & CNRS, LP2N UMR 5298, F-33400 Talence, France*
[3] *Centre de recherche CERVO, Québec, Canada*
[4] *Department of Psychiatry and Neuroscience, Université Laval, Québec, Canada*
[5] *Department of Physics, Freie Universität Berlin, Arnimallee 14, 14195 Berlin, Germany*
[6] *Institute of Organic Chemistry und Biochemistry, Freie Universität Berlin, Takustr. 3, 14195 Berlin, Germany*
[7] *Department of Chemistry, Faculty of Science, Lorestan University, Khorram Abad, Iran*
[†] *Present address: Universität des Saarlandes, Fachrichtung Physik, Campus E2.6, D-66123 Saarbrücken, Germany*
[*] *Correspondence should be addressed to laurent.cognet@u-bordeaux.fr*

**Abstract**

*The design of single-molecule photoswitchable emitters was the first milestone toward the advent of single-molecule localization microscopy that sets a new paradigm in the field of optical imaging. Several photoswitchable emitters have been developed but they all fluoresce in the visible or far-red ranges, missing the desirable near-infrared window where biological tissues are most transparent. Moreover, photocontrol of individual emitters in the near-infrared would be highly desirable for elementary optical molecular switches or information storage elements since most communication data transfer protocols are established in this spectral range. Here we introduce a novel type of hybrid nanomaterials consisting of single-wall carbon nanotubes covalently functionalized with photo-switching molecules that are used to control the intrinsic luminescence of the single nanotubes in the near-infrared (beyond 1 µm). We provide proof-of-concept of localization microscopy based on these bright photoswitchable near-infrared emitters.*

**Introduction**

Over the last decade, super-resolution microscopy revolutionized fluorescence microscopy by delivering optical images with resolutions below the diffraction limit, down to the nanometer scales. The majority of super-resolution approaches, such as stimulated emission depletion microscopy (STED)(*1*), photoactivated localization microscopy (PALM)(*2*) or stochastic optical reconstruction microscopy (STORM)(*3*) (and related techniques(*4*)), are based on the control of the emission properties of fluorescent molecules, in order to switch the emitters between on- and off-emission states. For instance, the advent of single molecule localization microscopy (*e.g.*, PALM, STORM), which is based on the super-localization of single molecules, was directly related to the conception of photoswitchable fluorescent emitters having photo-controllable blinking properties at the single molecule level(*5–7*). To access the optical window of biological tissues, *i.e.* where tissue scattering and absorption are minimal and thus should allow localization microscopy at depth within biological tissue, the use of near-infrared (NIR) nanoprobes, with emission wavelength $\lambda$ >1 µm, will be crucial(*8–10*). However, the available red-shifted dyes do not exhibit strong emission properties at the single molecule level and photoactivable emitters in this spectral range do not yet exist. The lack of appropriate emitters for implementing super-resolution techniques is even more detrimental in the NIR than in visible light: the limit of diffraction – an upper boundary for resolution in conventional far field microscopy – increases linearly with wavelength ($1.22 \cdot \lambda/(2NA)$; *e.g.* ~450 nm for $\lambda$ = 1065 nm with high NA objectives.

Single-wall carbon nanotubes (CNTs) display strong optical resonances in the NIR(*11*) and the discovery of their NIR luminescence(*12*) soon paved the way to their detection in living cells(*13*) and whole animals(*14, 15*). They display exceptional luminescence signal stabilities in aqueous environments (~tens of min(*16, 17*)), superior to most other fluorescent nano-probes used in single molecule experiments for biological applications. NIR luminescent CNTs have thus proven to be unparalleled single molecule probes due to their brightness and photostability, for intracellular single-molecule tracking in cultured cells(*18–20*). Using small carbon nanotubes ($L < 300\ nm$), it was previously shown that pointing accuracy of ~50 nm, which corresponds to ~$\lambda/20$, could be achieved(*17, 21*). Their emission spectral range was also key for long-term single molecule tracking at depth in living brain tissue to uncover the extracellular dimensions of the brain at nanometer resolutions. Blinking photoluminescent CNTs have already been observed in acidic environments or through charge transfer near surfaces which allowed achieving super-resolution imaging of CNTs emission sites(*21*) or quenching sites(*22, 23*). However, in these reports, photoblinking was observed without control of the blinking rate or efficiency. For future application of CNTs as NIR single molecules in localization microscopy, the first building block is thus still missing. Toward this aim, here we report the design, experimental characterization and modelling through simulations, of photoswitchable CNTs having controlled blinking properties in the NIR (1065 nm) at the single nanotube level.

**Results**

Our approach is based on Spiropyran-Merocyanine (SP-MC) molecules covalently attached to CNTs through a nitrene-based cycloaddition reaction (Fig. S1A)(*24*). The process noticeably preserves the conjugation of the sp$^2$ network and ensures the CNTs to remain fluorescent upon covalent functionalization (up to 4% density of functional groups; Fig. S2A). The unique functionalization yields a fully conjugated SP-CNT hybrid as demonstrated by the behavior of the π-electrons after conversion from SP to MC: In previous non-covalent approaches, the π-electron remained confined to the MC molecule, giving rise to the charge distribution that yields the strong MC dipole moment and its characteristic visible absorption band(*25*). In our covalent approach, there is no interruption of the conjugation between SP and nanotubes. The π-electron released by the switch after isomerization to MC conjugates over the extended CNT, uplifting the position of its Fermi level(*24*).

Illuminating a solution of SP-CNTs with a UV lamp induced a ~50% loss of luminescence intensity within a few seconds (Fig. S2A&B). This effect was previously shown to be fully reversible; it is due to the photoisomerization of SP-MC molecules which modulates charge transfer to the CNT(*24*). The excitons are insensitive to the presence of the functional group in the SP state, due to the π-preserving character of the functionalization. Upon UV illumination, the MC configuration is favored and the subsequent charge transfer from the MC to the CNT induces non-radiative recombination of the exciton (*i.e.* photoluminescence partially quenched; Fig. S2).

We will first reveal the origin of this incomplete loss of fluorescence upon UV illumination by performing a single molecule study. We prepared (10,2) CNTs carrying randomly distributed SP-MC molecules (~1 per 100 carbon atoms(*24*)). We imaged individual CNTs spin-coated on a microscope glass cover slip excited in wide-field configuration using a circularly polarized 730 nm laser line (Ti:Sa) for resonant excitation on the second order transition ($S_{22}$). Photoluminescence of the (10,2) CNTs which occurs at ~1065 nm was imaged using a near-infrared InGaAs camera (Fig. 1A&B and Methods section). Single (10,2) CNTs showed bright and stable photoluminescence when illuminated *via* the resonant $S_{22}$ excitation (Fig. 1C&D; black curves). We then added a lamp UV illumination (centered at 387 nm, with a bandwidth of 11 nm), and observed an overall decrease in intensity associated with independent blinking events (Fig. 1C; violet curves, Movie 1). This behavior is the origin for the incomplete loss of luminescence at the ensemble level. Indeed, averaging 79 nanotube intensity profiles, Fig. 1D, exhibits a dynamic that is similar to the isomerization observed through absorption spectroscopy on SP-CNTs in solution (Fig. S1C). The 40% loss in intensity in Fig. 1D is comparable to the ~50% reduction measured for the (10,2) nanotube on bulk SP-CNT samples (Fig. S2). We conclude that the photoblinking of the individual CNTs is responsible of the incomplete loss of photoluminescence observed at the ensemble level.

*Monte-Carlo simulation of the photoblinking behavior of single CNTs.* To better understand the blinking behavior of the nanotube under UV illumination, we developed a model for the complete photophysical processes in individual CNTs, taking into account the photogeneration of excitons, their spatial diffusion along the CNT backbones and their radiative or non-radiative recombination (Fig. 2A&B). This model bears similarities with previous work investigating how permanent quenching defects can affect the CNT luminescence properties(*26*) but here, the fast photoinduced quenching dynamics due to SP-MC isomerizations is critical and has to be additionally taken into account. We use the reported experimental value $l_d$ = 180 nm(*27*) for the exciton diffusion length in sodium cholate stabilized (10,2) CNTs and we consider that SP-MC molecules are randomly distributed along the nanotube. We further model the transition probability rate transfer statistics, between the SP and the MC conformations, by a 2-state Markov process. We assume that all SP-MC molecules follow the same rate transfer probability and independent transitions. We define $t_{SP}$ and $t_{MC}$ as the average residence times in the SP and the MC states, *i.e.* the time a molecule stays in the respective conformation, and the ratio between the residence times $\varphi = t_{SP}/t_{MC}$. Residence times should not be confused with the isomerization times required by the single SP-MC molecules to switch from one state to the other. The isomerization times occur on a much faster time scale (~hundreds of picoseconds) and are assumed to happen instantly. The generation rate of excitons is assumed to be constant along the nanotube. The probability for an exciton created at $x_0$ to recombine at position $x$ is given by $c(x,x_0,l_d) = 1/2l_d \cdot e^{-|x-x_0|/l_d}$. During its diffusion process, the exciton recombines non-radiatively if it encounters a molecule in the MC conformation or the end of the CNT (which also acts as a photoluminescence quencher(*28*)). Otherwise the exciton will recombine radiatively emitting a photon. Unless stated otherwise, the linear density of SP-MC molecules, $N_{SM}$, is one per nanometer of nanotube(*24*). Diffusion and recombination are repeated for every exciton generated, and the state of each SP-MC molecules is updated at the end of each integration time (50 ms to match experimental observations). Fig. 2C&D present examples of intensity time traces generated by this simulation considering a CNT length of $L$ = 300 nm, respectively for $\varphi$ = 100 with varying $t_{SP}$ and for $t_{SP}$ = 50 s with varying $\varphi$.

An overall glance at the time traces suggests that replicating the experimental data requires $t_{SP} \gg t_{MC}$. This can be explained by the fact that even if the large majority of the SP-MC molecules are in the SP configuration, the probability that a photogenerated exciton encounters an MC is high because of the high number of SP-MC molecules present on the nanotube along the exciton diffusion range ($l_d \sim \varphi/N_{SM}$). The SP-MC molecules thus act in a cooperative way to quench excitons. In contrast, $t_{MC} \sim t_{SP}$ (i.e. $l_d \gg \varphi/N_{SM}$) drastically increases the quenching probability during the exciton diffusion process resulting in full quenching instead of blinking.

From the simulations we identify two parameters that allow to determine $t_{SP}$ and $t_{MC}$ from the experimental measurement of the blinking behavior of individual CNTs: (i) the mean intensity ratio between the stable state (before UV illumination, all molecules in SP form) and the blinking state (UV illumination, mixture of SP and MC forms) and (ii) the temporal autocorrelation function (ACF) of the blinking time traces. The mean intensity ratio for a given CNT length increases with $\varphi$ and tends to 1 for large $\varphi$ (Fig. 3A). Interestingly, the intensity ratio is almost independent of $t_{SP}$ and depends mainly on $\varphi$ (Fig. S4). An analytical solution of the intensity ratio is also obtained (Methods) and displayed on Fig. 3A for $L = 300$ nm and varying $\varphi$. This shows that $\varphi$ can be analytically determined from the experimental intensity ratio knowing $L, l_d$ & $N_{SM}$ with an error $\frac{|\Delta\varphi|}{\varphi}$ less than 12% for $L = 300 \pm 50$ nm (Fig. 3A).

The autocorrelation decay time, $t_c$, increases with $t_{SP}$ for a given $\varphi$ and decreases with $\varphi$ for a given $t_{SP}$ (Fig. 3B). This can be explained by interpreting $t_{SP}$ as a scaling factor at a given $\varphi$. The plateaus in the intensity time traces increase with $t_{SP}$. Conversely, for constant $t_{SP}$ an increase in $\varphi$ induces faster varying events and decreases the autocorrelation time constant. From Fig. 3C, it can be observed that for a given nanotube length and linear defect density, the intensity ratio and the autocorrelation decay time fully describe the systems and provide estimates of the residence times of the SP and the MC states ($t_{SP}$ and $t_{MC}$).

Interestingly, the temporal ACFs of the photoluminescence during blinking events are directly linked to the linear density of quenching molecules and the residence times in SP and MC states (Methods). Temporal time traces can thus be simulated using our model for different sets of parameters varying $t_{SP}$ and $t_{MC}$ and consequently $\varphi$ (Fig. 3C&D). The diffusion length, $l_d$, couples the SP/MC states of adjacent molecules because the faith of the excitons depends on all quenching sites encountered along their path. For this reason, we find that the autocorrelation decay time $t_c < (t_{SP} \cdot t_{MC})/(t_{SP} + t_{MC})$ (Fig. 3D and inset). The temporal ACF of a single perfect fluorophore undergoing photo-intermittency between an "on" state and an "off" state (*e.g.* excited state and triplet state) would have given an autocorrelation decay time $t_c = (t_{on} \cdot t_{off})/(t_{on} + t_{off})$ (Methods).

*Analysis of the experimental blinking traces for the determinations of $t_{SP}$ and $t_{MC}$.* The simulations described above indicate that the intensity time traces of a single CNT (Fig. 4A) provide $\varphi$ using the ratio between the mean intensity when the nanotube is irradiated with UV light and the mean intensity before the UV irradiation. From the traces of 18 nanotubes we obtain a mean value of $\varphi = 80 \pm 18$ (Mean $\pm$ S.E.M) (Fig. 4A&C). We next calculate the temporal ACFs (Fig. 4B) of the intensity time traces of each nanotube and obtain a mean decay time constant $t_C = 0.5 \pm 0.1\ s$ (Mean $\pm$ S.E.M) (Fig. 4D). From the temporal autocorrelation time constants and using the slope obtained in the simulation data presented in Fig. 3D, we estimate the ratio $(t_{SP} \cdot t_{MC})/(t_{SP} + t_{MC}) = 0.4 \pm 0.1$ (Mean $\pm$ S.E.M). Combining with the knowledge of $\varphi = t_{SP}/t_{MC}$, this yields the values for $t_{SP} = 46 \pm 11\ s$ (Mean $\pm$ S.E.M) (Fig. 4E) and $t_{MC} = 0.6 \pm 0.1\ s$ (Mean $\pm$ S.E.M) (Fig. 4F) from the experimental data.

*Super-localization and super-resolution imaging of individual photoswitchable nanotubes.* From single nanotube temporal intensity profiles and using a forward-backward non-linear filtering technique(*29*), intensity plateaus are identified. Subtracting the average image of consecutive plateaus provides images of the single emitting sites. By fitting a two-dimensional Gaussian to these images, super-localization of individual blinking sites are obtained in a similar manner as used in super-resolution localization microscopy(*4*). The pointing accuracy of single-molecules can be estimated from(*30*). It depends on the strength of fluorescence signal ($S_i$), the background noise ($B_r$), the gain of the camera ($G = 20e^-/ADU$), the pixel size of the camera (a = 0.49 $\mu$m/pixel) and full-width half maximum of the fitted Gaussian ($\sigma$). The pointing accuracy is then given by $2\sqrt{2 \cdot Var_x}$ where $Var_x = \frac{\sigma_a^2}{s}\left(\frac{16}{9} + \frac{8\pi\sigma_a^2 b^2}{sa^2}\right)$, with $\sigma_a^2 = \sigma^2 + \frac{a^2}{12}$, $s = \frac{S_i}{G}$ and $b = \frac{B_r}{G}$. For this analysis, only plateaus longer than $t_C \approx 0.5\ s$ are considered and images from plateaus are averaged to increase the pointing accuracy. Blinking detections with >7700 intensity units are used to generate super-localization of individual CNTs in order to guarantee a pointing accuracy of $\omega_{FWHM} \sim 50$ nm corresponding to $\sim \lambda/20$. For display in Fig. 5A&B, single CNT detections are convolved with a two-dimensional Gaussian of $\omega_{FWHM} = 50$ nm to take into account the localization accuracy of the single emitters.

On occasion, owing to the blinking statistics of the photoswitchable CNTs, we can reveal the presence of distinct CNT segments that cannot be resolved initially. In Fig. 5A, different CNT segments are super-resolved 320 nm away from each other's corresponding to λ/3.3. These super-resolved images provide the proof-of-principle that photoswitchable CNTs will be suitable for localization super-resolution microscopy applications in the near-infrared.

**Conclusion**

In this work, we show that photoswitchable CNTs can be created by conjugating SP/MC molecules onto the CNTs *via* triazine linkers. Single nanotube experiments are presented to measure the blinking dynamics of the photoswitchable CNTs. Monte-Carlo simulations of 2-state Markov process taking into account the spatio-temporal exciton dynamics occurring in CNTs fully reproduced the experimental results. Combining simulations with the knowledge of $t_{SP}$ and $t_{MC}$, provides the means to create photo-induced blinking CNTs having arbitrary dynamics by varying the density of functionalization or illumination. It should thus be possible to tune these parameters in order to generate photoswitchable CNTs having blinking rates optimized for super-resolution imaging of densely labelled structures. The control of the emission properties of single CNTs using light is the first building block toward super-resolution studies in the NIR of biological samples(*4*) using CNTs. For this goal, future development will imply bioconjugation of the CNTs for specific labelling of cellular structures. Photoswitchable CNTs might also find applications in the field of information science as elementary optical molecular switches or information storage elements operating in the near-infrared.

**Materials and Methods**

*SP-CNT preparation and characterization.* HiPCO single-walled carbon nanotubes were purchased from Unidym (batch #SP0295, diameter 0.8-1.2 nm, median length 300 nm). 2,4,6-trichloro-1,3,5-triazine (cyanuric chloride or triazine), 2,3,3-trimethylindolenine, 5-nitrosalicylaldehyde were provided from Sigma-Aldrich. Sodium azide and N-methyl-2-pyrrolidone were purchased from Merck. Solvents and materials were used as they received and without further purifications. The synthesis of the SP-CNTs proceeded in two steps, following Ref.(*24*): Attachment of the triazine moieties onto the tubes (CNTs-Trz) and subsequent growth of the spiropyran/merocyanine switching moiety on the tubes (SP-CNTs).

*Synthesis of triazine-functionalized single-walled carbon nanotubes (CNTs-Trz):* CNTs (1 g) were dispersed in N-methyl-2-pyrrolidone (150 ml) and sonicated for 0.5-1 h. The dispersion was stirred for 1-2 h at 25 °C and cooled down to 0 °C. 2,4,6-1,3,5-trichloro-triazine (10 g, 54 mmol) was dissolved in N-methyl-2-pyrrolidone (50 ml) and the obtained solution was slowly added to the CNTs dispersion at 0 °C. Sodium azide (1.76 g, 27 mmol) was added to the mixture and stirred for 2 h at 0 °C followed by 12 h stirring at 70 °C for 12 h. The product was purified by centrifugation re-dispersed in water and different organic solvents (acetone, toluene, and chloroform), and lyophilized for storage and characterization. The triazine functionalization was characterized by elemental analysis, XPS spectroscopy, thermogravimetric analysis, Raman, and infrared spectroscopy, see Ref.(*24*) for details.

*Synthesis of spiropyran-functionalized CNTs (SP-CNT):* SP-CNTs were synthesized in a multi-step synthetic process(*24*). The indole segment was attached to the surface of CNTs-Trz by a nucleophilic reaction between the chlorine atoms of the triazine groups and 2,3,3-trimethylindolenine. CNT-Trz (0.2 g) was added to N-methyl-2-pyrrolidone (150 ml) and sonicated for 1 h. A solution of 2,3,3-trimethylindolenine in N-methyl-2-pyrrolidone (2 ml, 12.47 mmol/10ml) was added to this mixture at 0 °C and stirred for 1 h. After sonication at 25 °C, for 1 h, the temperature of reaction was raised to 65 °C and the mixture was stirred under nitrogen atmosphere for 4 days. The product (CNT-indole) was purified by centrifugation and re-dispersion in water and organic solvents such as acetone, chloroform, and THF. CNT-indole was changed to CNT-indolene by dispersing it in a saturated aqueous solution of NaOH and sonicated for 30 min. The mixture was stirred at room temperature for 5 h and then purified by repeated washing with water and centrifugation. SP-CNT was synthesized by adding 5-nitrosalicylaldehyde (2.5 g, 1.19 mmol) to a well-sonicated and degassed dispersion of CNT-indolene (0.1 g) in dry ethanol (70 ml) at 25 ºC. After sonication at 25 °C with 35 kHz for 2 h and stirring at 70 ºC for 12 h, the solvent was evaporated and the mixture was re-dispersed in ethanol, chloroform, water, toluene, and acetone and collected by centrifugation at 5000 rpm for 5 min. CNT functionalization by SP was confirmed by elemental analysis, XPS spectroscopy, Raman scattering, UV/Vis, infrared and photoluminescence spectroscopy as described in detail in Ref.(*24*), see also Fig. S2.

*Switching of SP-CNTs in solution and preparation of single SP-CNT samples.* To monitor the switching behavior of the hybrids in suspension we prepare a bulk solution by dissolving the SP-CNTs in water (density of tubes 0.127 g/L) and adding sodium cholate (1 wt %). After tip-sonication (Bandelin Sonopuls HD 2070, 1 h at 16 W) and centrifugation (Hettich Mikro 220 R centrifuge, 30000 g for 1 h), we collected the supernatant for optical characterization. 2d excitation-emission spectroscopy of the bulk solution was performed with a Nanolog spectrofluorometer from Horiba (Xenon lamp source and liquid-Nitrogen cooled InGaAs detector). Kinetic absorption measurements of the bulk suspensions under ultraviolet illumination were performed with a spectrophotometer from Thermo-Fisher coupled with a handheld ultraviolet lamp emitting at 365 nm as ultraviolet light source.

*Near-infrared single nanotube imaging.* Single nanotube photoluminescence imaging was performed with an inverted microscope equipped with a 1.40 NA 60x objective. A volume of 10 μL drop of SP-CNT suspension was spin-coated on polyvinylpyrrolidone (PVP) coated glass coverslips in order to have isolated nanotubes. The excitation source consisted of a tunable Ti:Sa laser emitting at a wavelength of 730 nm to preferentially excite (10,2) CNTs at the resonance excitation on the second order transition ($S_{22}$). The excitation intensity was kept at 10 kW/cm$^2$ with circularly polarized light. CNTs were detected by an InGaAs camera (Xenics Xeva 1.7 320 TE3) at 20 frames per second with a pixel size of 0.49 $\mu$m. A band-pass filter Z1064/10x (Chroma) was used to detect the (10,2) CNT-emitted fluorescence. An epifluorescence white-light excitation illumination with FF01-387/11 (Semrock) were used for the UV illumination.

*Modelling of the mean photoluminescence intensity ratio during blinking events.* To model the mean photoluminescence intensity probability during the blinking processes (normalized to the luminescence without defects), we derive an analytical expression for the mean intensity of a segment of arbitrary length $X$. It corresponds to the probability of an exciton recombining before reaching the end of the nanotube and gives: $p(x_0, l_d, X) = \int_0^X c(x, x_0, l_d)dx / \int_{-\infty}^{+\infty} c(x, x_0, l_d)dx =$ $1 - 1/2\left(e^{-(X-x_0)/l_d} + e^{-x_0/l_d}\right)$. Assuming that excitons are generated uniformly along the nanotube segment, allows to estimate the average intensity of a nanotube segment of arbitrary length $X$ by integrating $p(x_0, l_d, X)$ over the nanotube segment: $int(l_d, X) = \int_0^X p(x_0, l_d, X) dx_0 = X + l_d \cdot \left(e^{-X/l_d} - 1\right)$ (Equation (1)). The integrated intensity from a nanotube with $n$ quenching sites is given by the sum of the intensity of each $(n+1)$ nanotubes segments. The intensity of each segment is given by equation (1) and Fig. S3A. The mean normalized intensity for a nanotube of length $L = 300$ nm can then be calculated (Fig. S3B). This curve was obtained by numerically generating 50000 random configurations having $n$ quenchers. The relation is well approximated by $Int(L, l_d, n) = 1/(1 + a(l_d/L) \cdot n)$ where $a(l_d/L)$ depends only on the length ratio $l_d/L$ and relates the probability of an exciton to encounter a quencher ($a(l_d/L) = c_1 \cdot (l_d/L)^{c_2}/(1 + c_3 \cdot (l_d/L)^{c_2})$, where $c_1, c_2$ and $c_3$ are constants) (Fig. S3C). The average luminescence intensity of a nanotube of length $L$ with diffusion length $l_d$ having $N_{SM}$ SP-MC molecules is then obtained by weighing each $int(L, l_d, n)$ by the probability of having exactly $n$ SP-MC molecules in the MC state on the nanotube at a given time: $Poi(N_{SM}/\varphi, n) = (N_{SM}/\varphi)^n \cdot e^{-N_{SM}/\varphi}/n!$. Where $N_{SM}/\varphi$ is the average number of $n$ SP-MC molecules in the MC state and $\varphi = t_{SP}/t_{MC}$. The final normalized intensity is then given by $int(L, l_d, N_{SM}, \varphi) = \sum_{n=0}^{\infty} 1/(1 + a(L, l_d) \cdot n) \cdot Poi(N_{SM}/\varphi, n)$ (Equation (2)) and displayed on Fig. 3A for $L = 300 \pm 50$ nm and varying $\varphi$. From equation (2), the ratio $\varphi$ can thus be determined knowing $L, l_d$ & $N_{SM}$ with an error $|\Delta\varphi|/\varphi < 12\%$ for $L = 300 \pm 50$ nm Error on $|\Delta\varphi|/\varphi < 13\%$ for $N_{SM} = 1 \pm 0.1$ per nm & $|\Delta\varphi|/\varphi < 10\%$ for $l_d = 200 \pm 50$ nm (Fig. 3A & Fig. S5).

*Monte-Carlo simulations of temporal luminescence intensity profiles.* Nanotubes of length $L$ with SP-MC linear density of $N_{SM}$ are simulated. The SP-MC molecules are randomly distributed along the nanotube length. The positions of the SP-MC molecules are fixed for each simulated nanotube. Each of them was simulated to undergo transition between two states SP and MC with residence times $t_{SP}$ and $t_{MC}$, respectively. Initially all SP-MC molecules are in the SP conformation. The generation probability for excitons is assumed uniform along the nanotube and the probability for the exciton, created at $x_0$, to recombine at a position $x$ is given by $c(x, x_0, l_d) = 1/2l_d \cdot e^{-|x-x_0|/l_d}$, where $l_d$ is the diffusion length of the excitons. If the exciton encounters a molecule in the MC conformation or the end of the tube during the diffusion process it recombines non-radiatively; otherwise it recombines in a radiative way and emits a photon. This process is repeated for every exciton generated and the state of each SP-MC molecules was updated at the end of each integration time. The exciton generation rate is set to 133 exciton nm$^{-1}$ s$^{-1}$. A minimum of 5 nanotubes was simulated for each set of parameters ($L, l_d, N_{SM}, t_{SP}$ and $t_{MC}$).

*Temporal autocorrelation of the intensity time trace*. The temporal statistics of a single fluorophore undergoing photo-intermittency between an "on" and "off" state (*e.g.* excited state and triplet state) corresponds to a two-states Markovian model. Temporal autocorrelation function (ACF) of the fluorophore emission defines an autocorrelation decay time given by $t_c = (t_{on} \cdot t_{off})/(t_{on} + t_{off})$, where $1/t_{on}$ is the decay rate from the ''on'' state to the ''off'' state and $1/t_{off}$ is the decay rate from the ''off'' state to the ''on'' state. Having more than one fluorophore undergoing the same blinking statistics will not modify $t_c$ assuming their transitions from one state to another are independent. Using our simulation, we also investigated the role of each parameters on $t_c$. As expected, decreasing the number of quenching sites that each exciton encounters increases $t_c$ up to the limit of $(t_{SP} \cdot t_{MC})/(t_{SP} + t_{MC})$ (*e.g.* ↓ $l_d$, ↓ $N_{SM}$ or ↑ $\varphi$) (Fig. 3D inset).

**Acknowledgments**

This work was supported by CNRS, the Agence Nationale de la Recherche (ANR-14-OHRI-0001-01, ANR-16-CE29-0011-03), IdEx Bordeaux (ANR-10-IDEX-03-02), Conseil Régional d'Aquitaine (2011-1603009) and the France-BioImaging national infrastructure (ANR-10-INBS-04-01). A.G.G. acknowledges financial support from the Fondation pour la Recherche Médicale and the Fonds Recherche du Québec–Nature et Technologies.

# Figures

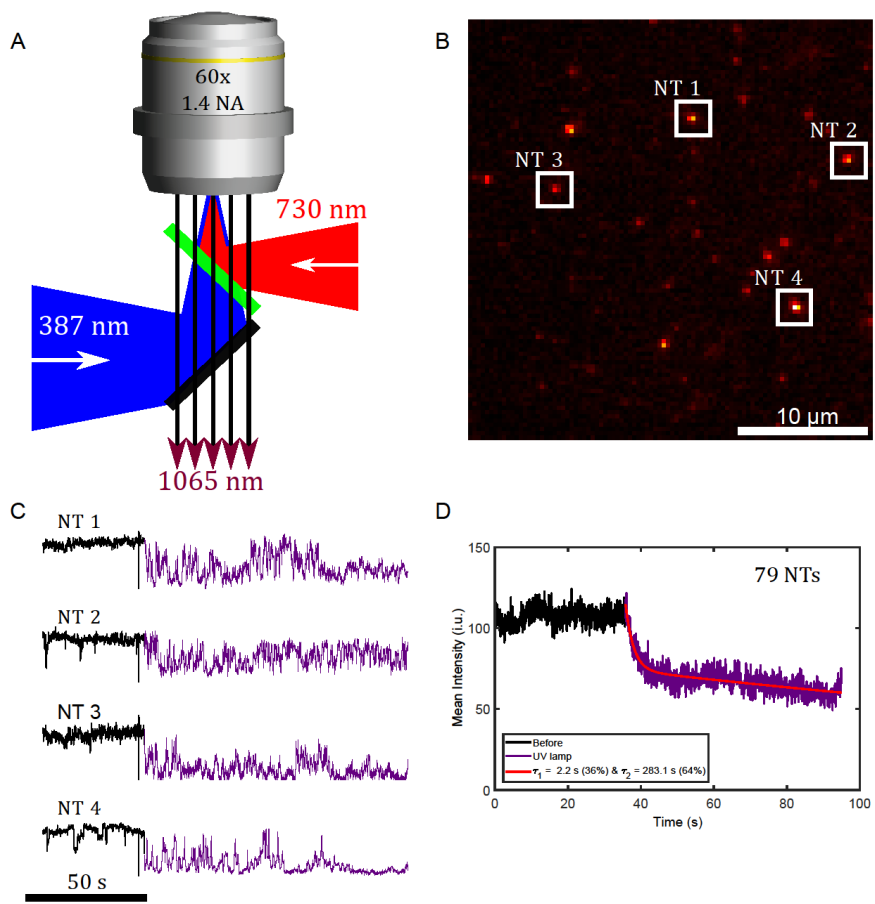

**Fig. 1.** *UV irradiation induces luminescence intermittency on carbon nanotubes functionalized with SP-MC.* (**A**) Schema representing the excitation and emission fluorescence set-up with UV illumination. (**B**) Wide-field NIR image of (10,2) carbon nanotubes spin-coated on PVP illuminated with a 730 nm circular polarized laser (~10 kW/cm$^2$). Scale bar = 10 μm. (**C**) Luminescence time traces showing that UV irradiation induces emission intermittency and reduces the mean intensity. Imaging rate = 20 Hz and temporal scale bar = 50 s. (**D**) Time evolution of the average of 79 carbon nanotubes before and during UV irradiation.

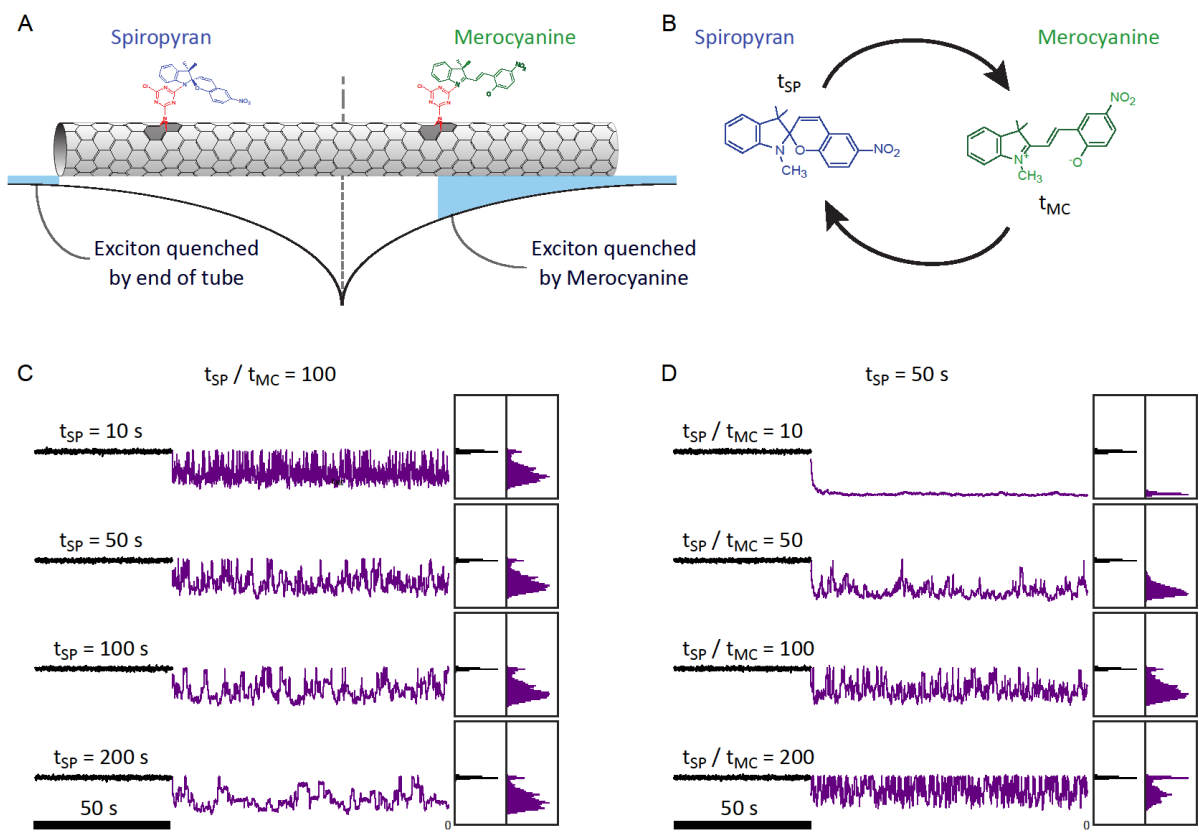

**Fig 2.** *Simulation modelling of the excitonic photophysical processes generating the luminescence.* (**A**) Schematic representation of the nanotube functionalized with Spiropyran-Merocyanine molecules. An exciton generated in the middle of the nanotube and its probability of recombining is shown. Only excitons that recombine before encountering a Merocyanine molecule or the end of the nanotube emit a photon. (**B**) The Spiropyran-Merocyanine group corresponds to a two-state model with transition rates defined as $a_{SP} = 1/t_{SP}$ and $a_{MC} = 1/t_{MC}$. Simulated time traces are presented for $\varphi = a_{MC}/a_{SP} = 100$ with varying $t_{SP}$ (**C**) and for $t_{MC} = 50$ s with varying $\varphi$ (**D**). Simulated measurement rate = 20 Hz. Nanotube length and diffusion lengths were $L = 300$ nm, $l_d = 180\ nm$ and $N_{SM} = 1$ per nm. The exciton generation rate was 133 exciton nm$^{-1}$ s$^{-1}$.

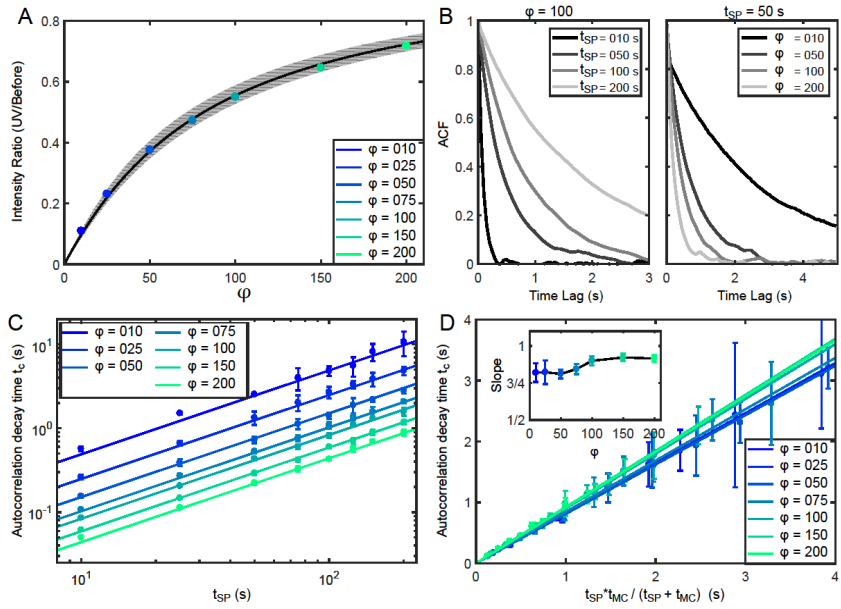

**Fig 3.** *Characterization of the luminescence temporal time traces.* (**A**) Graph showing the influence of the ratio $\varphi = a_{MC}/a_{SP}$ for $t_{MC} = 50$ s and $l_d = 180$ nm on the ratio of the average luminescence intensities before and during the UV irradiation. The grey shaded area corresponds to the error on the estimation of $\varphi$ for $L = 300 \pm 50$ nm. (**B**) Examples of temporal autocorrelation functions (ACFs) for the time traces presented in Fig. 2C&D for $\varphi = 100$ with varying $t_{SP}$ and for $t_{MC} = 50$ s with varying $\varphi$. (**C**) Graph summarizing the impact of varying $\varphi$ and $t_{SP}$ on the temporal autocorrelation decay constant. For Fig. 3A-C, $L = 300$ nm, $l_d = 180$ nm and $N_{SM} = 1$ per nm. (**D**) Graph summarizing the impact of varying $\varphi$ and $t_{SP}$ on the temporal autocorrelation decay constant plotted *vs.* $t_c^{ind} = (t_{SP} \cdot t_{MC})/(t_{SP} + t_{MC})$. The slope $t_c = slope \cdot t_c^{ind}$ is also shown in the inset as a function of $\varphi$.

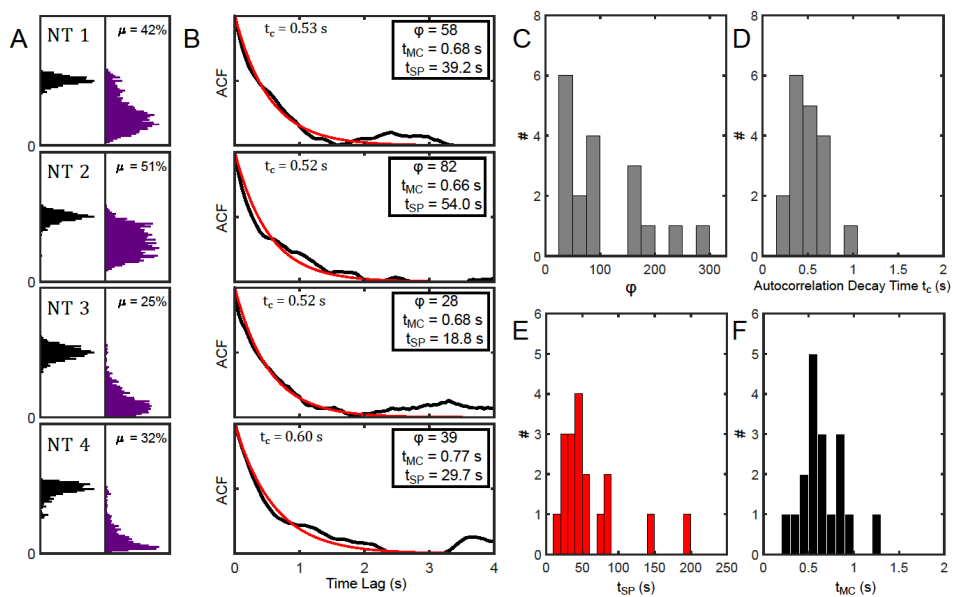

**Fig 4.** *Estimation of the switching rate dynamics of the Spiropyran-Merocyanine molecules from the intensity time trace of individual carbon nanotubes.* (**A**) Four normalized intensity histograms and the corresponding temporal autocorrelation functions (**B**) for the carbon nanotubes shown in Fig. 1. (**C**) and (**D**) represent the histograms of the intensity ratio $\varphi$ and $t_c$, respectively, obtained from individual carbon nanotube luminescence time traces. Knowing $\varphi$ and $t_c$, the values of $t_{SP}$ (**E**) and $t_{MC}$ (**F**) can be estimated. Values correspond to Mean ± S.E.M.

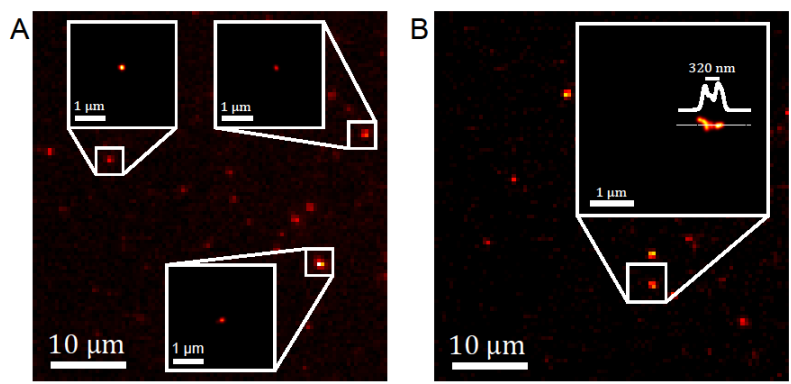

**Fig 5.** *Super-localization and super-resolution imaging of single carbon nanotubes using photo-controlled luminescence intermittency.* Wide-field NIR zoomed images of (10,2) carbon nanotubes spin-coated on PVP illuminated with a 730 nm circular polarized laser (~10 kW/cm$^2$). The zoomed regions of interests correspond to reconstructed super-localization of individual carbon nanotubes (**A**) and super-resolved image of closely located nanotubes (**B**) using intensity transitions from all the blinking steps in the acquired movies (20 Hz). The super-resolved image shows different nanotube segments ~ 320 nm apart that could not be resolved in the wide-field image. For display, each localization is convolved with a Gaussian having a $\omega_{FWHM}$ = 50 nm. Scale bar = 10 μm.

# Supplementary figures

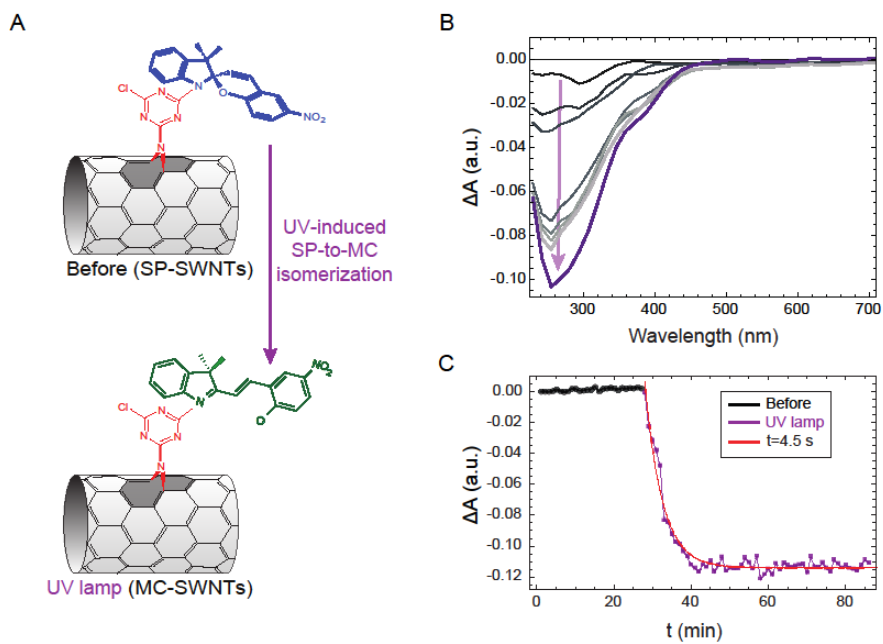

**Fig S1.** *Spiropyran/merocyanine nanotube hybrid.* (**A**) Spiropyran is covalently attached to a carbon nanotube *via* a triazine anchor group. The functionalization preserves the conjugated π-electron system of the CNT. Upon UV illumination the SP functional group (top) transforms into MC (bottom). (**B**) This process is monitored by the appearance of an optical absorption band in the UV. (**C**) Dynamics of the SP-MC isomerization follows a mono-exponential decay with a time constant of 4.5 s.

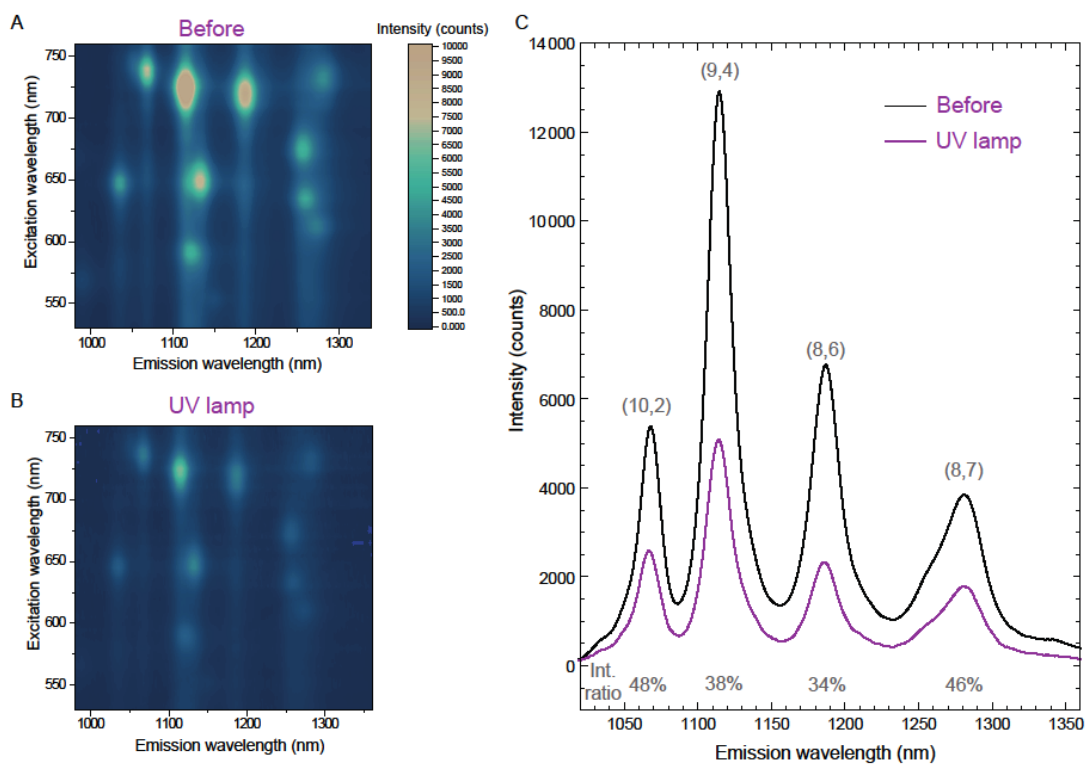

**Fig S2.** *SP-CNT luminescence in solution.* (**A**) Intensity of the CNT luminescence as a function of excitation and emission wavelength for the SP-CNT hybrid. Each peak corresponds to a particular CNT chirality (*n,m*). (**B**) Upon UV illumination the hybrid switches into the MC-CNT form, which is accompanied by an overall loss in intensity. (**C**) Luminescence before (black line) and after (purple) UV illumination excited with 730 nm excitation wavelength (identical to the single tube experiments). The nanotube chiralities and their loss in luminescence intensity (intensity ratio before/after UV) are indicated. The (10,2) tube shows an intensity ratio of 48%.

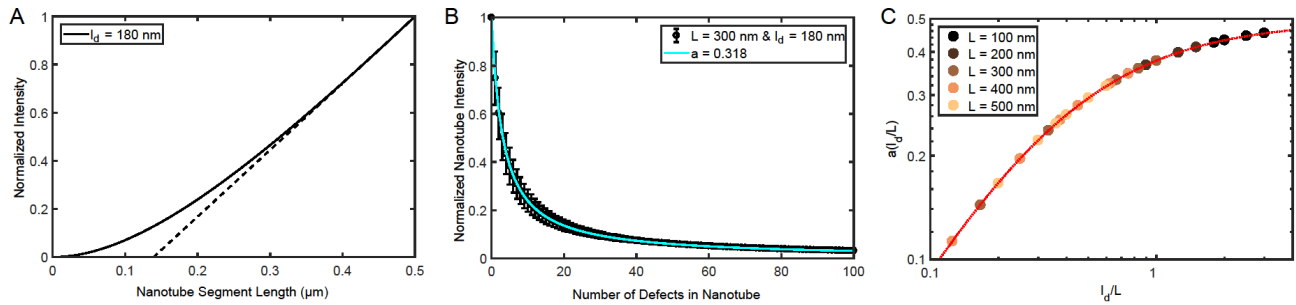

**Fig S3.** *Theoretical modelling of the luminescent intensity of a nanotube with quencher molecules.* (**A**) Luminescence intensity of a perfect nanotube segment of length $x$ with $l_d = 180$ nm. This curve was calculated using equation (1) and the intensity was normalized by the intensity of a nanotube of length $L = 500$ nm. To generate this graph, the illumination intensity was assumed constant. (**B**) The influence on the whole nanotube intensity ($L = 300$ nm and $l_d = 180$ nm) when $n$ molecules undergo a transition from a luminescent and to a quenching state. The intensity can be approximated by the function $1/(1 + a \cdot n)$ where the variable $a$ is only a function of the ratio $l_d/L$ (**C**). The red dotted line corresponds to the best fit of the equation $a(l_d/L) = c_1 \cdot (l_d/L)^{c_2}/(1 + c_3 \cdot (l_d/L)^{c_2})$, where $c_1, c_2$ and $c_3$ are constants.

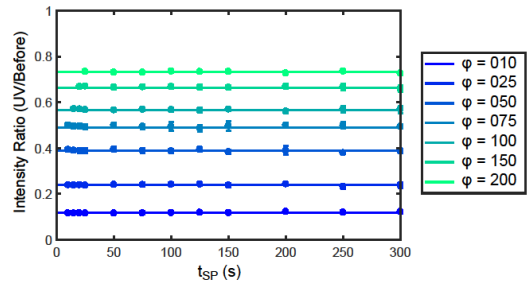

**Fig S4.** *Influence of the residence times on the intensity ratio.* Graph showing that the luminescence intensity ratio during and before UV irradiation is dependent on the residence times only through their ratio $\varphi = t_{SP}/t_{MC}$, thus enabling unambiguous determination of this parameter from the measured mean intensities. This data set is obtained for $N_{SM} = 1$ per nm, $L = 300$ nm and $l_d = 180$ nm and was used to generate Fig. 3A.

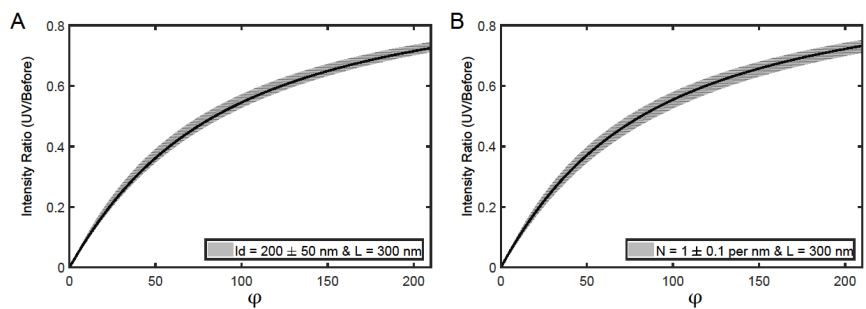

**Fig S5.** *Error estimation of the parameter Theoretical $\varphi$.* The influence of $\varphi$ on the mean intensity ratio is presented for a nanotube segment of $L = 300$ nm. The grey shaded areas correspond to the error of the estimated $\varphi$ for $N_{SM} = 1.0 \pm 0.2$ per nm, $L = 300$ nm & $l_d = 180$ nm (**A**) and $N_{SM} = 1$ per nm, $L = 300$ nm & $l_d = 200 \pm 50$ nm (**B**).